\PassOptionsToPackage{bookmarks=false}{hyperref}
\documentclass[fleqn,10pt]{wlpeerj}

\usepackage{booktabs}
\usepackage{enumitem}
\usepackage{dblfloatfix}
\usepackage{multirow}
\usepackage{footnote}
\usepackage{array}
\usepackage{amsmath}
\usepackage{listings}
\usepackage{color}
\usepackage{colortbl}
\usepackage{balance}
\usepackage{rotating}
\usepackage{tablefootnote}
\usepackage{ifthen}
\usepackage{tikz}
\usepackage{xspace}
\usepackage{soul}
\usepackage[export]{adjustbox}
\usepackage{url}

\setitemize{leftmargin=*}

\newif\ifIsDoubleBlind
\IsDoubleBlindfalse

\definecolor{ColorWeakLightBlue}{RGB}{220, 240, 255}
\definecolor{ColorHighlightLightBlue}{RGB}{191, 225, 255}
\definecolor{ColorLightGray}{RGB}{220, 220, 220}
\newcommand{\tableRowBgColor}{ColorWeakLightBlue}

\newcolumntype{L}[1]{>{\raggedright\let\newline\\\arraybackslash\hspace{0pt}}m{#1}}
\newcolumntype{R}[1]{>{\raggedleft\let\newline\\\arraybackslash\hspace{0pt}}m{#1}}
\newcommand{\halfBaselineskip}{0.25\baselineskip}
\newcommand{\miniSkip}{0.15\baselineskip}

\newcommand{\formatMetricIndex}[1]{M#1}
\newcommand{\formatItemName}[1]{\mbox{\textit{#1}}}
\newcommand{\formatItemNameSmall}[1]{\formatItemName{#1}}
\newcommand{\formatCaptionDetails}[1]{\textmd{\small{#1}}}
\newcommand{\formatResultBox}[1]{\vspace{\miniSkip} \noindent \fbox{\parbox{\linewidth}{#1}} \vspace{\miniSkip}}

\newcommand{\coloredSquare}[1]{\tikz\filldraw[#1,fill=#1] (0,0) rectangle ++(4pt,4pt);}
\newcommand{\coloredLine}[1]{\textcolor{#1}{\rule[0.5ex]{0.32cm}{1.2pt}}}

\renewcommand{\formatCaptionDetails}[1]{#1}
\renewcommand{\formatItemNameSmall}[1]{\footnotesize{\formatItemName{#1}}}

\renewcommand{\halfBaselineskip}{0.5\baselineskip}
\renewcommand{\miniSkip}{0.3\baselineskip}

\renewcommand{\formatResultBox}[1]{\begin{center} \noindent \fbox{\parbox{13cm}{#1}} \end{center}}

\let\originalCite\cite
\renewcommand{\cite}[1]{[\originalCite{#1}]}

\begin{document}

\ifIsDoubleBlind
	\newcommand{\citeResults}{\cite{Niedermayr2018T4DataDoubleBlind}}
\else
	\newcommand{\citeResults}{\cite{Niedermayr2018T4Data}}
\fi

\newcommand{\minSuppVal}{10\%}
\newcommand{\minConfVal}{90\%}
\newcommand{\fdr}{fault-density reduction\xspace}
\newcommand{\fdrShort}{factor\xspace}
\newcommand{\fdrLong}{fault-density reduction factor\xspace}

\newcommand{\paperTitle}{Too Trivial To Test?}
\newcommand{\paperSubTitle}{An Inverse View on Defect Prediction to Identify Methods with Low Fault Risk}


\title{\paperTitle \\ {\paperSubTitle}}

\author[1]{Rainer Niedermayr}
\author[2]{Tobias R\"ohm}
\author[3]{Stefan Wagner}

\affil[1,2]{CQSE GmbH, Garching b. M\"unchen, Germany}
\affil[1,3]{University of Stuttgart, Stuttgart, Germany} 

\corrauthor[1]{Rainer Niedermayr}{niedermayr@cqse.eu}


\begin{abstract}
\textbf{\textit{Background.}}
Test resources are usually limited and therefore it is often not possible to completely test an application before a release.
To cope with the problem of scarce resources,
 development teams can apply defect prediction to identify fault-prone code regions.
However, defect prediction tends to low precision in cross-project prediction scenarios.
\textbf{\textit{Aims.}}
We take an inverse view on defect prediction
 and aim to identify methods that can be deferred when testing
 because they contain hardly any faults due to their code being ``trivial''.
We expect that characteristics of such methods might be project-independent,
 so that our approach could improve cross-project predictions.
\textbf{\textit{Method.}}
We compute code metrics and apply association rule mining to create rules for identifying methods with low fault risk.
We conduct an empirical study to assess our approach with six Java open-source projects containing precise fault data at the method level.
\textbf{\textit{Results.}}
Our results show
 that inverse defect prediction can identify approx. 32--44\% of the methods of a project to have a low fault risk;
 on average, they are about six times less likely to contain a fault than other methods.
In cross-project predictions with larger, more diversified training sets, 
 identified methods are even eleven times less likely to contain a fault.
\textbf{\textit{Conclusions.}}
Inverse defect prediction supports the efficient allocation of test resources
 by identifying methods that can be treated with less priority in testing activities
 and is well applicable in cross-project prediction scenarios.
\end{abstract}

\flushbottom
\maketitle
\thispagestyle{empty}

\section{Introduction}
\label{Sec:Introduction}

In a perfect world, it would be possible to \textit{completely} test every new version of a software application before it was deployed into production.
In practice, however, software development teams often face a problem of scarce test resources.
Developers are busy implementing features and bug fixes,
 and may lack time to develop enough automated unit tests to comprehensively test new code~\cite{ostrand2005predicting, menzies2004good}.
Furthermore, testing is costly and, depending on the criticality of a system, it may not be cost-effective to expend equal test effort to all components~\cite{zhang2007predicting}.
Hence, development teams need to prioritize and limit their testing scope by restricting the code regions to be tested~\cite{menzies2003can, bertolino2007software}.
To cope with the problem of scarce test resources, development teams aim to test code regions
 that have the best cost-benefit ratio regarding fault detection.
To support development teams in this activity, defect prediction has been developed and studied extensively in the last decades~\cite{hall2012systematic, d2012evaluating, catal2011software}.
Defect prediction identifies code regions that are likely to contain a fault and should therefore be tested~\cite{menzies2007data, Weyuker2008FaultPrediction}.


This paper suggests, implements, and evaluates another view on defect prediction: inverse defect prediction (IDP).
The idea behind IDP is to identify code artifacts (e.g., methods) that are so \textit{trivial}
 that they contain hardly any faults and thus can be deferred or ignored in testing.
Like traditional defect prediction,
 IDP also uses a set of metrics that characterize artifacts,
 applies transformations to pre-process metrics,
 and uses a machine-learning classifier to build a prediction model.
The difference rather lies in the predicted classes.
While defect prediction classifies an artifact either as \textit{buggy or non-buggy},
 IDP identifies methods that exhibit a \textit{low fault risk} (LFR) with high certainty
 and does not make an assumption about the remaining methods,
 for which the fault risk is at least medium or cannot be reliably determined.
As a consequence, the objective of the prediction also differs.
Defect prediction aims to achieve
 a high recall, such that as many faults as possible can be detected,
 and a high precision, such that only few false positives occur.
In contrast, IDP aims to achieve high precision to ensure that low-fault-risk methods contain indeed hardly any faults,
 but it does not necessarily seek to predict all non-faulty methods.
Still, IDP needs to achieve a certain recall such that a reasonable reduction potential arises
 when treating LFR methods with a lower priority in QA activities.

\vspace{\halfBaselineskip}
\textbf{Research goal:} We want to study whether IDP can reliably identify code regions that exhibit only a low fault risk,
 whether ignoring such code regions---as done silently in defect prediction---is a good idea,
 and whether IDP can be used in cross-project predictions.

\vspace{\halfBaselineskip}
To implement IDP, we calculated code metrics for each method of a code base and trained a classifier for methods with low fault risk using association rule mining.
To evaluate IDP, we performed an empirical study with the Defects4J dataset~\cite{just2014defects4j} consisting of real faults from six open-source projects.
We applied static code analysis and classifier learning on these code bases and evaluated the results.
We hypothesize that IDP can be used to pragmatically address the problem of scarce test resources.
More specifically, we hypothesize that a generalized IDP model can be used to identify code regions that can be deferred when writing automated tests if none yet exist, as is the situation for many legacy code bases.

\vspace{\halfBaselineskip}
\textbf{Contributions:}
1)~The idea of an inverse view on defect prediction:
While defect prediction has been studied extensively in the last decades, it has always been employed to identify code regions with \textit{high} fault risk.
To the best of our knowledge, the present paper is the first to study the identification of code regions with \textit{low} fault risk explicitly.
2)~An empirical study about the performance of IDP on real open-source code bases.
3)~An extension to the Defects4J dataset~\cite{just2014defects4j}:
To improve data quality and enable further research---reproduction in particular---we provide
 code metrics for all methods in the code bases and an indication whether they were changed in a bug-fix patch,
 a list of methods that changed in bug fixes only to preserve API compatibility,
 and association rules to identify low-fault-risk methods.

\vspace{\halfBaselineskip}
The remainder of this paper is organized as follows.
Section~\ref{Sec:Background} provides background information about association rule mining.
Section~\ref{Sec:Related_Work} discusses related work.
Section~\ref{Sec:Approach} describes the IDP approach, i.e.,
 the computation of the metrics for each method,
 the data pre-processing,
 and the association rule mining to identify methods with low fault risk.
Afterwards, Section~\ref{Sec:Empirical_Study} summarizes the design and results of the IDP study with the Defects4J dataset.
Then, Section~\ref{Sec:Discussion} discusses the study's results, implications, and threats to validity.
Finally, Section~\ref{Sec:Conclusion} summarizes the main findings and sketches future work.
\section{Association Rule Mining}
\label{Sec:Background}

Association rule mining is a technique for identifying relations between variables in a large dataset
 and was introduced by Agrawal et al. in 1993~\cite{agrawal1993mining}.
A dataset contains \textit{transactions} consisting of a set of \textit{items} that are binary attributes.
An \textit{association rule} represents a logical implication of the form
	$\{$~\textit{antecedent}~$\}$ $\rightarrow$ $\{$~\textit{consequent}~$\}$
 and expresses that the \textit{consequent} is likely to apply if the \textit{antecedent} applies.
Antecedent and consequent both consist of a set of items and are disjoint.
The \textit{support} of a rule expresses the proportion of the transactions that contain both antecedent and consequent out of all transactions.
It is related to the significance of the itemset~\cite{simon2011simple}.
The \textit{confidence} of a rule expresses the proportion of the transactions that contain both antecedent and consequent
 out of all transactions that contain the antecedent.
It can be considered as the precision~\cite{simon2011simple}.
A rule is \textit{redundant} if a more general rule with the same or a higher confidence value exists~\cite{bayardo1999constraint}.

Association Rule Mining has been successfully applied in defect prediction studies~\cite{
 song2006software, czibula2014software, ma2010software, zafar2012finding}.
A major advantage of association rule mining is the natural comprehensibility of the rules~\cite{simon2011simple}. 
Other commonly used machine-learning algorithms for defect prediction,
 such as support vector machines (SVM) or Naive Bayes classifiers,
 generate black-box models, which lack interpretability.
Even decision trees can be difficult to interpret due to the subtree-replication problem~\cite{simon2011simple}.
Another advantage of association rule mining is that the gained rules implicitly extract high-order interactions among the predictors.
\section{Related Work}
\label{Sec:Related_Work}
Defect prediction is an important research area that has been extensively studied~\cite{hall2012systematic, catal2009systematic}.
Defect prediction models use code metrics~\cite{menzies2007data, nagappan2006mining, d2012evaluating, zimmermann2007predicting},
 change metrics~\cite{nagappan2005use, hassan2009predicting, kim2007predicting},
 or a variety of further metrics (such as code ownership~\cite{bird2011don, rahman2011ownership}, developer interactions~\cite{meneely2008predicting, lee2011micro}, dependencies to binaries~\cite{zimmermann2008predicting}, mutants~\cite{bowes2016mutation}, code smells~\cite{palomba2016smells}) to predict code areas that are especially defect-prone.
Such models allow software engineers to focus quality-assurance efforts on these areas
 and thereby support a more efficient resource allocation~\cite{menzies2007data, Weyuker2008FaultPrediction}.

Defect prediction is usually performed at the component, package or file level~\cite{nagappan2005use, nagappan2006mining, bacchelli2010popular, scanniello2013class}.
Recently, more fine-grained prediction models have been proposed to narrow down the scope for quality-assurance activities.
Kim et al. presented a model to classify software changes~\cite{kim2008classifying}.
Hata et al. applied defect prediction at the method level
 and showed that fine-grained prediction outperforms coarse-grained prediction at the file or package level if efforts to find the faults are considered~\cite{hata2012bug}.
Giger et al. also investigated prediction models at the method level~\cite{giger2012method}
 and concluded that a Random Forest model operating on change metrics can achieve good performance.
More recently, Pascarella et al. replicated this study and confirmed the results~\cite{pascarella2018re}.
However, they reported that a more realistic inter-release evaluation of the models shows a dramatic drop in performance with results close to that of a random classifier
 and concluded that method-level bug prediction is still an open challenge~\cite{pascarella2018re}.
It is considered difficult to achieve sufficiently good data quality at the method level~\cite{hata2012bug, shippey2016esemData};
 publicly available datasets have been provided in \cite{shippey2016esemData}, \cite{just2014defects4j}, and \cite{giger2012method}.

Cross-project defect prediction predicts defects in projects for which no historical data exists by using models trained on data of other projects~\cite{zimmermann2009cross, xia2016hydra}.
He et al. investigated the usability of cross-project defect prediction~\cite{he2012investigation}.
They reported that cross-project defect prediction works only in few cases and requires careful selection of training data.
Zimmermann et al. also provided empirical evidence that cross-project prediction is a serious problem~\cite{zimmermann2009cross}.
They stated that projects in the same domain cannot be used to build accurate prediction models
 without quantifying, understanding, and evaluating process, data and domain.
Similar findings were obtained by Turhan et al., who investigated the use of cross-company data for building prediction models~\cite{turhan2009relative}.
They found that models using cross-company data can only be ``useful in extreme cases such as mission-critical projects, where the cost of false alarms can be afforded''
 and suggested using within-company data if available.
While some recent studies reported advances in cross-project defect prediction~\cite{xia2016hydra, zhang2016cross, xu2018cross},
 it is still considered as a challenging task.

Our work differs from the above-mentioned work in the target setting:
 we do not predict artifacts that are fault-prone, but instead identify artifacts (methods) that are very unlikely to contain any faults.
While defect prediction aims to detect as many faults as possible (without too many false positives), and thus strives for a high recall~\cite{mende2009revisiting},
 our IDP approach strives to identify those methods that are not fault-prone to a high certainty.
Therefore, we optimized our approach towards the precision in detecting low-fault-risk methods and considered the recall as less important.
To the best of our knowledge, this is the first work to study low-fault-risk methods.
Moreover, as far as we know, cross-project prediction has not yet been applied at the method level.
To perform the classification, we applied association rule mining.
Association rule mining has previously been applied with success in defect prediction~\cite{song2006software, morisaki2007defect, czibula2014software, ma2010software, karthik2010defect, zafar2012finding}.

\section{IDP Approach}
\label{Sec:Approach}

\begin{figure}
	\centering
	\includegraphics[width=0.8\linewidth,clip]{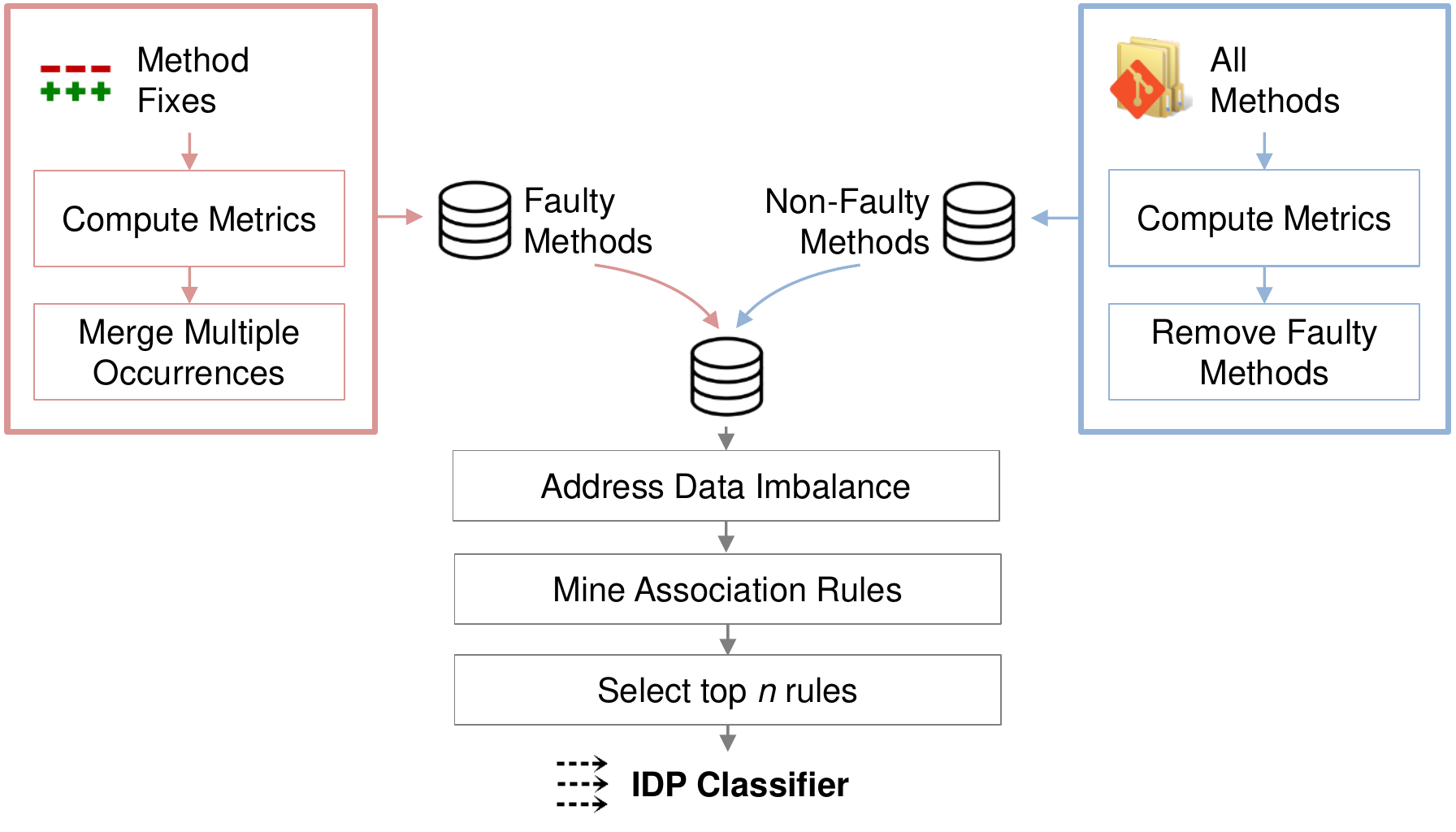}
	\caption{Overview of the approach.
					 \formatCaptionDetails{Metrics for faulty methods are computed at the faulty state;
					 metrics for non-faulty methods are computed at the state of the last bug-fix commit.}
	}
	\label{Fig:ApproachOverview}
\end{figure}

This section describes the inverse defect prediction approach, which identifies low-fault-risk (LFR) methods.
The approach comprises the computation of source-code metrics for each method, the data pre-processing before the mining, and the association rule mining.
Figure~\ref{Fig:ApproachOverview} illustrates the steps.

\subsection{Metric Computation}
\label{Sec:Metrics}

\begin{table}
	\centering
	\caption{Computed metrics for each method.}
	\begin{tabular}{rll}
		\toprule
														 & Metric Name & Type \\
		\midrule
			 \formatMetricIndex{1} & Source Lines of Code (SLOC) & length \\
			 \formatMetricIndex{2} & Cyclomatic Complexity (CC) & complexity \\
			 \formatMetricIndex{3} & Max. Nesting Depth & max. value \\
			 \formatMetricIndex{4} & Max. Method Chaining & max. value \\
			 \formatMetricIndex{5} & Unique Variable Identifiers & unique count \\
		\midrule
			 \formatMetricIndex{6} & Anonymous Class Declarations & count \\
			 \formatMetricIndex{7} & Arithmetic In- or Decrementations & count \\
			 \formatMetricIndex{8} & Arithmetic Infix Operations & count \\
			 \formatMetricIndex{9} & Array Accesses & count \\
			\formatMetricIndex{10} & Array Creations & count \\
			\formatMetricIndex{11} & Assignments & count \\
			\formatMetricIndex{12} & Boolean Operators & count \\
			\formatMetricIndex{13} & Cast Expressions & count \\
			\formatMetricIndex{14} & Catch Clauses & count \\
			\formatMetricIndex{15} & Comparison Operators & count \\
			\formatMetricIndex{16} & If Conditions & count \\
			\formatMetricIndex{17} & Inner Method Declarations & count \\
			\formatMetricIndex{18} & Instance-of Checks & count \\
			\formatMetricIndex{19} & Instantiations & count \\
			\formatMetricIndex{20} & Loops & count \\
			\formatMetricIndex{21} & Method Invocations & count \\
			\formatMetricIndex{22} & Null Checks & count \\
			\formatMetricIndex{23} & Null Literals & count \\
			\formatMetricIndex{24} & Return Statements & count \\
			\formatMetricIndex{25} & String Literals & count \\
			\formatMetricIndex{26} & Super-Method Invocations & count \\
			\formatMetricIndex{27} & Switch-Case Blocks & count \\
			\formatMetricIndex{28} & Synchronized Blocks & count \\
			\formatMetricIndex{29} & Ternary Operations & count \\
			\formatMetricIndex{30} & Throw Statements & count \\
			\formatMetricIndex{31} & Try Blocks & count \\
	  \midrule
			\formatMetricIndex{32} & All Conditions & count \\
			\formatMetricIndex{33} & All Arithmetic Operations & count \\
		\midrule
			\formatMetricIndex{34} & Is Constructor & boolean \\
			\formatMetricIndex{35} & Is Setter & boolean \\
			\formatMetricIndex{36} & Is Getter & boolean \\
			\formatMetricIndex{37} & Is Empty Method & boolean \\
			\formatMetricIndex{38} & Is Delegation Method & boolean \\
			\formatMetricIndex{39} & Is ToString Method & boolean \\
		\bottomrule
	\end{tabular}
	\label{Tbl:Metric_Set}
\end{table}
Like defect prediction models, IDP uses metrics to train a classifier for identifying low-fault-risk methods.
For each method, we compute the source-code metrics listed in Table~\ref{Tbl:Metric_Set}
 that we considered relevant to judge whether a method is trivial.
They comprise established length and complexity metrics used in defect prediction,
 metrics regarding occurrences of programming-language constructs,
 and categories describing the purpose of a method.

\textit{SLOC} is the number of source lines of code, i.e., LOC without empty lines and comments.
\textit{Cyclomatic Complexity (CC)} corresponds to the metric proposed by McCabe~\cite{mcCabe1976complexity}.
Despite this metric being controversial~\cite{shepperd1988critique, hummel2014mcCabe}---due
 to the fact that it is not actionable, difficult to interpret, and high values do not necessarily translate to low readability---it
 is commonly used as variable in defect prediction~\cite{menzies2004assessing, zimmermann2007predicting, menzies2002metrics}.
Furthermore, a low number of paths through a method could be relevant for identifying low-fault-risk methods.
\textit{Maximum nesting depth} corresponds to the ``maximum number of encapsulated scopes inside the body of the method''~\cite{ndepend2017nesting}.
Deeply nested code is more difficult to understand, therefore, it could be more fault-prone.
\textit{Maximum method chaining} expresses the maximum number of chain elements of a method invocation.
We consider a method call to be chained if it is directly invoked on the result from the previous method invocation.
The value for a method is
 zero if it does not contain any method invocations,
 one if no method invocation is chained,
 or otherwise the maximum number of chain elements
 (e.g., two for \texttt{getId().toString()}, three for \texttt{getId().toString().subString(1)}).
\textit{Unique variable identifiers} counts the distinct names of variables that are used within the method.
The following metrics, metrics \formatMetricIndex{6} to \formatMetricIndex{31}, count the occurrences of the respective Java language construct~\cite{gosling2013java}.

Next, we derive further metrics from the existing ones.
They are redundant, but correlated metrics do not have any negative effects on association rule mining (except on the computation time)
 and may improve the results for the following reason:
 if an item generated from a metric is not frequent,
 rules with this item will be discarded because they cannot achieve the minimum support;
 however, an item for a more general metric may be more frequent and survive.
The derived metrics are:
\begin{itemize}[noitemsep]
	\item \textit{All Conditions}, which sums up \textit{If Conditions}, \textit{Switch-Case Blocks}, and \textit{Ternary Operations}
	 (\formatMetricIndex{16} + \formatMetricIndex{27} + \formatMetricIndex{29})
	\item \textit{All Arithmetic Operations}, which sums up \textit{Incrementations}, \textit{Decrementations}, and \textit{Arithmetic Infix Operations}
	 (\formatMetricIndex{7} + \formatMetricIndex{8})
\end{itemize}

Furthermore, we compute to which of the following categories a method belongs (a method can belong to zero, one, or more categories):
\begin{itemize}[noitemsep]
	\item \textit{Constructors:} Special methods that create and initialize an instance of a class.
				They might be less fault-prone because they often only set class variables or delegate to another constructor.
	\item \textit{Getters:} Methods that return a class variable.
				They usually consist of a single statement and can be generated by the IDE.
	\item \textit{Setters:} Methods that set the value of a class variable.
				They usually consist of a single statement and can be generated by the IDE.
	\item \textit{Empty Methods:} Non-abstract methods without any statements.
			  They often exist to meet an implemented interface,
				 or because the default logic is to do nothing and is supposed to be overridden in certain sub-classes.
	\item \textit{Delegation Methods:} Methods that delegate the call to another method with the same name and further parameters.
				They often do not contain any logic besides the delegation.
	\item \textit{ToString Methods:} Implementations of Java's \texttt{toString} method.
				They are often used only for debugging purposes and can be generated by the IDE.
\end{itemize}

Note that we only use source-code metrics and do not consider process metrics.
This is because we want to identify methods that exhibit a low fault risk due to their \textit{code.}

Association rule mining computes frequent itemsets from categorical attributes;
 therefore, our next step is to discretize the numerical metrics.
(In defect prediction, discretization is also applied to the metrics:
 Shivaji et al.~\cite{shivaji2013reducing} and McCallum et al.~\cite{mccallum1998comparison} reported
 that binary values can yield better results than using counts when the number of features is low.)
We discretize as follows:

\begin{itemize}[noitemsep]
	\item For each of the metrics \formatMetricIndex{1} to \formatMetricIndex{5}, we inspect their distribution and create three classes.
				The first class is for metric values until the first tertile, the second class for values until the second tertile,
				and the third class for the remaining values.
  \item For all count metrics (including the derived ones), we create a binary ``has-no''-metric,
				which is true if the value is zero, e.g.,
				\mbox{\formatItemName{CountLoops} $= 0\ \ \ \Rightarrow\ \ \ $ \formatItemName{NoLoops} $=$ \texttt{true}}.
  \item For the method categories (setter, getter, \dots), no transformation is necessary as they are already binary.
\end{itemize}
\subsection{Data Pre-Processing}
\label{Sec:PreProcessData}
At this point, we assume that we have a list of faulty methods with their metrics at the faulty state (the list may contain a method multiple times if it was fixed multiple times)
 and a list of all methods.
Faulty methods can be obtained by identifying methods that were changed in bug-fix commits~\cite{zimmermann2007predicting, giger2012method, shippey2016esemData};
 we describe in Section~\ref{Sec:FaultDataExtraction} how we extracted faulty methods from the Defects4J dataset.

Prior to applying the mining algorithm, we have 
 1) to address faulty methods with multiple occurrences,
 2) to create a unified list of faulty and non-faulty methods,
 and 3) to tackle dataset imbalance.

\vspace{\halfBaselineskip}
1) A method may be fixed multiple times; in this case, a method appears multiple times in the list of the faulty methods.
However, each method should have the same weight and should therefore be considered only once.
Consequently, we consolidate multiple occurrences of the same method:
 we replace all occurrences by a new instance and apply majority voting to aggregate the binary metric values.
It is common practice in defect prediction to have a single instance of every method with a flag that indicates whether a method was faulty at least once~\cite{
 menzies2010defect, giger2012method, shippey2016esemData, mende2009revisiting}.

\vspace{\halfBaselineskip}
2) To create a unified dataset, we take the list of all methods,
 remove those methods that exist in the set of the faulty methods,
 and add the set of the faulty methods with the metrics computed \textit{at the faulty state.}
After doing that, we end up with a list containing each method exactly once and a flag indicating whether a method was faulty or not.

\vspace{\halfBaselineskip}
3) Defect datasets are often highly imbalanced~\cite{khoshgoftaar2010attribute}, with faulty methods being underrepresented.
Therefore, we apply \textit{SMOTE}\footnote{Synthetic Minority Over-sampling Technique}, a well-known algorithm for over- and under-sampling, to address imbalance in the dataset used for training~\cite{longadge2013class, chawla2002smote}.
It artificially generates new entries of the minority class using the nearest neighbors of these cases and reduces entries from the majority class~\cite{r2010dmrw}.
If we do not apply \textit{SMOTE} to highly imbalanced datasets,
  many non-expressive rules will be generated when most methods are not faulty.
For example, if 95\% of the methods are not faulty and 90\% of them contain a method invocation,
 rules with high support will be generated that use this association to identify non-faulty methods.
Balancing avoids those nonsense rules.
\subsection{IDP Classifier}
\label{Sec:Approach_Mining}
To identify low-fault-risk methods, we compute association rules of the type
 $\{$\formatItemName{Metric1}, \formatItemName{Metric2}, \formatItemName{Metric3}, \dots$\} \rightarrow \{$\formatItemName{NotFaulty}$\}$.
Examples for the metrics are \formatItemName{SlocLowestThird}, \formatItemName{NoNullChecks}, \formatItemName{IsSetter}.
A method that satisfies all metric predicates of a rule is not faulty to the certainty expressed by the confidence of the rule.
The support of the rule expresses how many methods with these characteristics exist,
 and thus, it shows how generalizable the rule is.

After computing the rules on a training set, we remove redundant ones (see Section~\ref{Sec:Background}) and order the remaining rules first descending by their confidence and then by their support.
To build the low-fault-risk classifier,
 we combine the top \textit{n} association rules with the highest confidence values using the logical-or operator.
Hence, we consider a method to have a low fault risk if at least one of the top \textit{n} rules matches.
To determine \textit{n},
 we compute the maximum number of rules until the faulty methods in the low-fault-risk methods exceed a certain threshold in the training set.

Of course, IDP can also be used with other machine-learning algorithms.
We decided to use association rule mining because of the natural comprehensibility of the rules (see Section~\ref{Sec:Background})
 and because we achieved a better performance compared to models we trained using Random Forest.
\section{Empirical Study}
\label{Sec:Empirical_Study}
This section reports on the empirical study that we conducted to evaluate the inverse defect prediction approach.

\newcommand{\rqNumFaultsInLfr}{RQ~1\xspace}
\newcommand{\rqNumSizeLfr}{RQ~2\xspace}
\newcommand{\rqNumCrossPrj}{RQ~3\xspace}

\newcommand{\rqTextFaultsInLfr}{\rqNumFaultsInLfr: How many faults do methods classified as ``low fault risk'' contain?}
\newcommand{\rqTextSizeLfr}{\rqNumSizeLfr: How large is the fraction of the code base consisting of methods classified as ``low fault risk''?}
\newcommand{\rqTextCrossPrj}{\rqNumCrossPrj: Is a trained classifier for methods with low fault risk generalizable to other projects?}

\subsection{Research Questions}
We investigate the following questions
 to research how well methods that contain hardly any faults can be identified
 and to study whether IDP is applicable in cross-project scenarios.

\vspace{\halfBaselineskip}
\textbf{\rqTextFaultsInLfr}
To evaluate the precision of the classifier, we investigate how many methods that are classified as ``low-fault-risk''
 (due to the triviality of their code)
 are faulty.
If we want to use the low-fault-risk classifier for determining methods that require less focus
 during quality assurance (QA) activities, such as testing and code reviews,
 we need to be sure that these methods contain hardly any faults.

\vspace{\halfBaselineskip}
\textbf{\rqTextSizeLfr}
We study how common low-fault-risk methods are in code bases
 to find out how much code is of lower importance for quality-assurance activities.
We want to determine which savings potential can arise if these methods are excluded from QA.

\vspace{\halfBaselineskip}
\textbf{\rqTextCrossPrj}
Cross-project defect prediction is used to predict faults in (new) projects, for which no historical fault data exists, by using models trained on other projects.
It is considered a challenging task in defect prediction~\cite{he2012investigation, zimmermann2009cross, turhan2009relative}.
As we expect that the characteristics of low-fault-risk methods might be project-independent,
 IDP could be applicable in a cross-project scenario.
Therefore, we investigate how generalizable our IDP classifier is for cross-project use.
\subsection{Study Objects}
\label{Sec:Study_Objects}
For our analysis, we used data from Defects4J, which was created by Just et al.~\cite{just2014defects4j}.
Defects4J is a database and analysis framework that provides real faults for six real-world open-source projects written in Java.
For each fault, the original commit before the bug fix (faulty version), the original commit after the bug fix (fixed version),
 and a minimal patch of the bug fix are provided.
The patch is minimal such that it contains only code changes that
 1) fix the fault
 and 2) are necessary to keep the code compilable
 (e.g., when a bug fix involves method-signature changes).
It does not contain changes that do not influence the semantics (e.g., changes in comments, local renamings),
 and changes that were included in the bug-fix commit but are not related to the actual fault (e.g., refactorings).
Due to the manual analysis, this dataset at the method level is much more precise than other datasets at the same level,
 such as \cite{shippey2016esemData} and~\cite{giger2012method},
 which were generated from version control systems and issue trackers without further manual filtering.
The authors of~\cite{just2014defects4j} confirmed that they considered every bug fix within a given time span.

Table~\ref{Tbl:Study_Objects} presents the study objects and their characteristics.
We computed the metrics \textit{SLOC} and \textit{\#Methods} for the code revision at the last bug-fix commit of each project;
 the numbers do not comprise sample and test code.
\textit{\#Faulty methods} corresponds to the number of faulty methods derived from the dataset.

\begin{table}
	\centering
	\caption{Study objects.
	}
	\begin{tabular}{lrrr}
		\toprule
		 Name 																		&	SLOC   & \#Methods & \#Faulty Meth. \\
		\midrule
		 JFreeChart	\textit{(Chart)}							&  81.6k &  6.8k &  39 \\  
		 Google \textit{Closure} Compiler					& 166.7k & 13.0k & 148 \\ 
		 Apache Commons \textit{Lang}							&  16.6k &  2.0k &  73 \\ 
		 Apache Commons \textit{Math}							&   9.5k &  1.2k & 132 \\ 
		 \textit{Mockito}													&  28.3k &  2.5k &  64 \\ 
		 Joda \textit{Time}												&  89.0k & 10.1k &  45 \\ 
		\bottomrule
	\end{tabular}
	\label{Tbl:Study_Objects}
\end{table}

\subsection{Fault Data Extraction}
\label{Sec:FaultDataExtraction}
\newcommand{\commitHashOrigFaulty}{\textbf{\textcolor[RGB]{192, 80, 77}{\texttt{f81f3f}}}}
\newcommand{\commitHashOrigFixed}{\textbf{\textcolor[RGB]{79, 129, 189}{\texttt{c1e8ed}}}}
\newcommand{\commitHashOurFaulty}{\textbf{\textcolor[RGB]{192, 80, 77}{\texttt{fa30f1}}}}

\begin{figure}
	\centering
	\includegraphics[width=0.7\linewidth,clip]{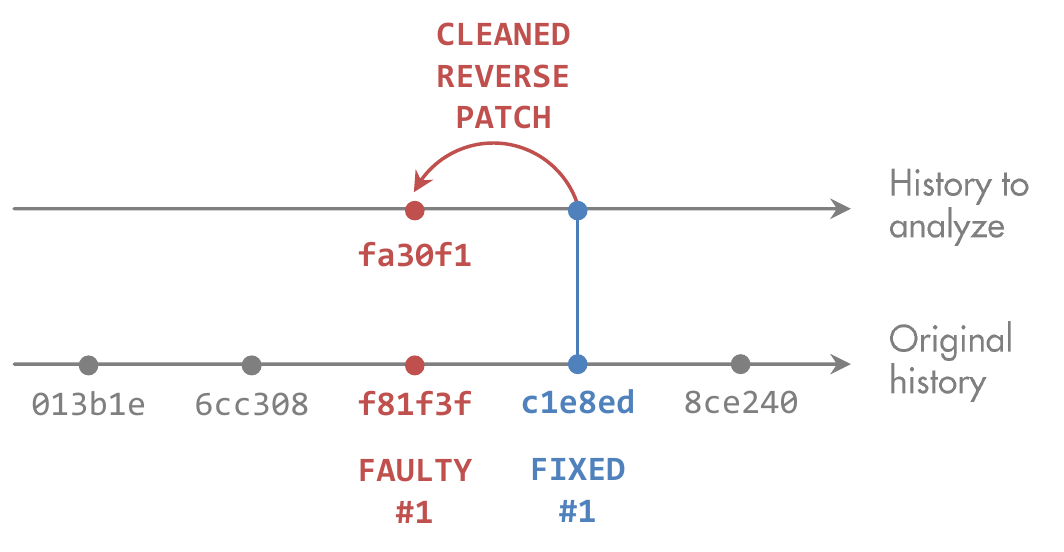}
	\caption{Derivation of faulty methods.
					\formatCaptionDetails{
					 The original bug-fix commit \commitHashOrigFixed{} to fix the faulty version \commitHashOrigFaulty{} may contain unrelated changes.
					 Defect4J provides a reverse patch, which contains only the actual fix.
					 We applied it to the fixed version \commitHashOrigFixed{} to get to \commitHashOurFaulty.
					 We then identified methods that were touched by the patch and computed their metrics at state \commitHashOurFaulty.}
					}
	\label{Fig:FaultyMethods}
\end{figure}
\begin{figure}
	\centering
	\includegraphics[width=0.85\linewidth,left]{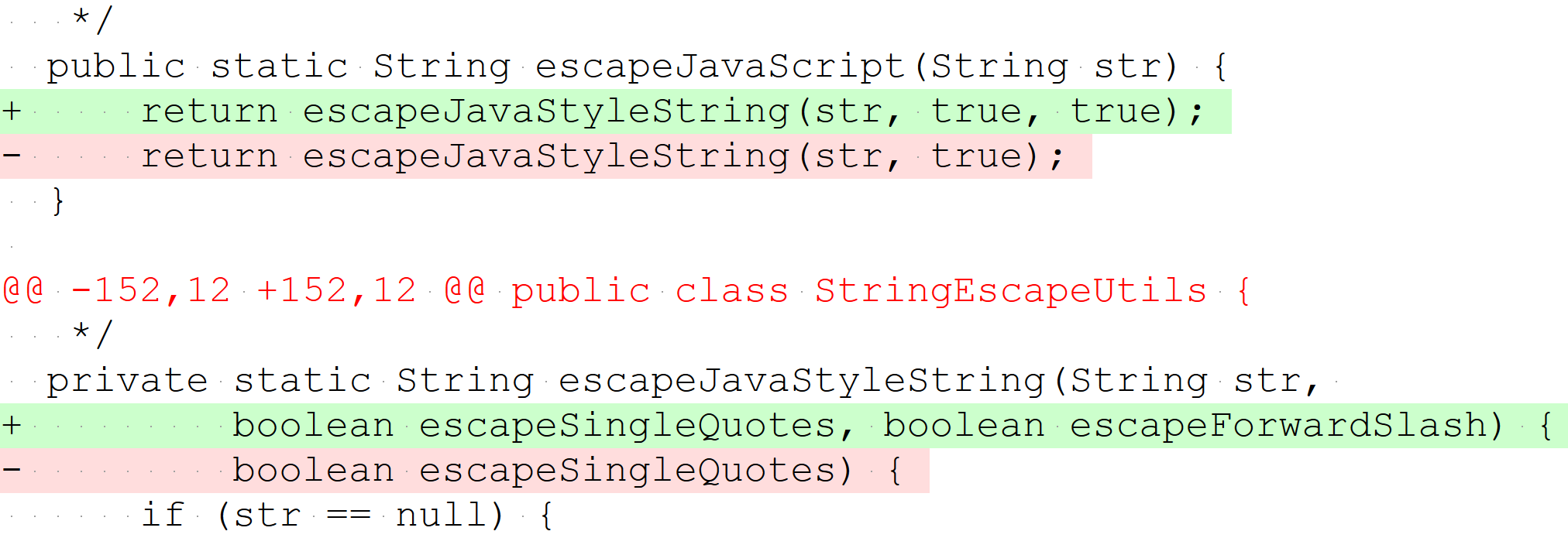}
	\caption{Example of method change without behavior modification to preserve API compatibility.
	\formatCaptionDetails{
	  The method \texttt{escapeJavaScript(String)} invokes \texttt{escapeJavaStyleString(String, boolean, boolean)}.
	  A further parameter was added to the invoked method;
	   therefore, it was necessary to adjust the invocation in \texttt{escapeJavaScript(String)}.
	  For invocations with the parameter value \texttt{true}, the behavior does not change [\textit{Lang}, patch~46, simplified].}}
	\label{Fig:DelegationMethod}
\end{figure}

Defects4J provides for each project a set of reverse patches\footnote{A reverse patch reverts previous changes.},
 which represent bug fixes.
To obtain the list of methods that were at least once faulty,
 we conducted the following steps for each patch.
First, we checked out the source code from the project repository at the original bug-fix commit
 and stored it as \textit{fixed version}.
Second, we applied the reverse patch to the fixed version to get to the code before the bug fix
 and stored the resulting \textit{faulty version}.

Next, we analyzed the two versions created for every patch.
For each file that was changed between the faulty and the fixed version,
 we parsed the source code to identify the methods.
We then mapped the code changes to the methods to determine which methods were touched in the bug fix.
After that, we had the list of faulty methods.
Figure~\ref{Fig:FaultyMethods} summarizes these steps.

We inspected all 395 bug-fix patches and found that 10 method changes in the patches do not represent bug fixes.
While the patches are minimal, such that they contain only bug-related changes (see Section~\ref{Sec:Study_Objects}),
 these ten method changes are semantic-preserving, only necessary because of changed signatures of other methods in the patch,
 and therefore included in Defects4J to keep the code compilable.
Figure~\ref{Fig:DelegationMethod} presents an example.
Although these methods are part of the bug fix, they were not changed semantically and do not represent faulty methods.
Therefore, we decided to remove them from the faulty methods in our analysis.
The names of these ten methods are provided in the dataset to this paper~\citeResults.
\subsection{Procedure}
\label{Sec:Procedure}

After extracting the faulty methods from the dataset, we computed the metrics listed in Section~\ref{Sec:Approach}.
We computed them for all faulty methods at their faulty version and for all methods of the application code\footnote{
  code without sample and test code}
 at the state of the fixed version of the last patch.
We used Eclipse JDT AST\footnote{\url{http://www.eclipse.org/jdt/}} to create an AST visitor for computing the metrics.
For all further processing, we used the statistical computing software R\footnote{\url{https://cran.r-project.org/}}.

\begin{figure}
	\centering
	\includegraphics[width=1.0\linewidth]{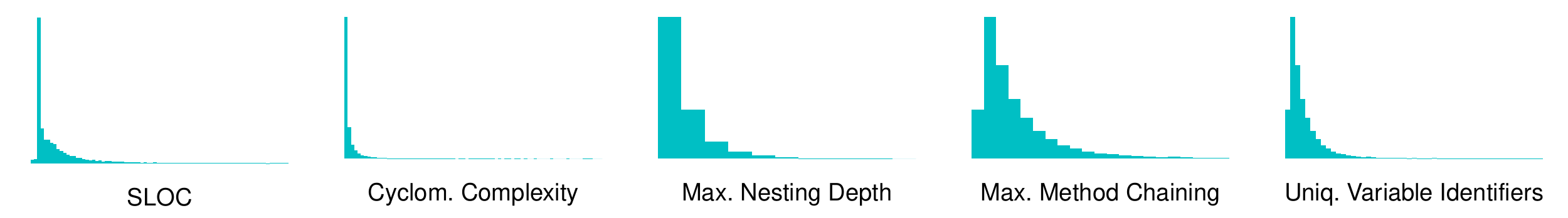}
	\caption{Metrics \formatMetricIndex{1} to \formatMetricIndex{5} are not normally distributed.}
	\label{Fig:MetricDistribution}
\end{figure}

To discretize the metrics \formatMetricIndex{1} to \formatMetricIndex{5}, we first computed their value distribution.
Figure~\ref{Fig:MetricDistribution} shows that their values are not normally distributed (most values are very small).
To create three classes for each of these metrics,\footnote{
 We did not use the \texttt{ntile} function to create classes,
	because it always generates classes of the same size,
	such that instances with the same value may end up in different classes
	(e.g., if 50\% of the methods have the complexity value 1, the first 33.3\% will end up in class 1,
	and the remaining 16.7\% with the same value will end up in class 2).
 }
 we sorted the metric values, and computed the values at the end of the first and at the end of the second third.
We then put 
 all methods until the last occurrence of the value at the end of the first third into class 1,
 all methods until the last occurrence of the value at the end of the second third into class 2,
 and all other methods into class 3.
Table~\ref{Tbl:Metric_Classes} presents the value ranges of the resulting classes.
The classes are the same for all six projects.


\begin{table}
	\centering
	\caption{Generated classes and their value ranges.}
	\begin{tabular}{lccc}
		\toprule
			Metric 											& Class 1 & Class 2 & Class 3 		 \\
		\midrule
			SLOC 												& $[0;3]$ & $[4;8]$ & $[9;\infty)$ \\
			Cyclomatic Complexity 			& $[1;1]$ & $[2;2]$ & $[3;\infty)$ \\
			Max. Nesting Depth					& $[0;0]$ & $[1;1]$ & $[2;\infty)$ \\
			Max. Method Chaining 				& $[0;1]$ & $[2;2]$ & $[3;\infty)$ \\
			Uniq. Variable Identifiers 	& $[0;1]$ & $[2;3]$ & $[4;\infty)$ \\
		\bottomrule
	\end{tabular}
	\label{Tbl:Metric_Classes}
\end{table}

We then aggregated multiple faulty occurrences of the same method (this occurs if a method is changed in more than one bug-fix patch)
 and created a unified dataset of faulty and non-faulty methods (see Section~\ref{Sec:PreProcessData}).

Next, we split the dataset into a training and a test set.
For \rqNumFaultsInLfr and \rqNumSizeLfr, we used 10-fold cross-validation~[\originalCite[Chapter~5]{witten2016data}].
Using the \textit{caret} package~\cite{caret2017doc},
 we randomly sampled the dataset of each project into ten stratified partitions of equal sizes.
Each partition is used once for testing the classifier, which is trained on the remaining nine partitions.
To compute the association rules for \rqNumCrossPrj---in which we study how generalizable the classifier is---for each project,
 we used the methods of the other five projects as training set for the classifier.

Before computing association rules,
 we applied the SMOTE algorithm from the \textit{DMwR} package~\cite{r2010dmrw}
 with a 100\% over-sampling and a 200\% under-sampling rate to each training set.
After that, each training set was equally balanced (50\% faulty methods, 50\% non-faulty methods).\footnote{
 We computed the results for the empirical study once with and once without addressing the data imbalance in the training set.
 The prediction performance was better when applying SMOTE, therefore, we decided to use it.}

We then used the implementation of the \textit{Apriori} algorithm~\cite{agrawal1994fast}
 in the \textit{arules} package~\cite{hahsler2017arules, hahsler2005arules}
 to compute association rules with \formatItemName{NotFaulty} as target item (rule consequent).
We set the threshold for the minimum support to \minSuppVal{}
 and the threshold for the minimum confidence to \minConfVal{} (support and confidence are explained in Section~\ref{Sec:Background}).
We experimented with different thresholds and these values produced good results
 (results for other configurations are in the dataset provided with this paper~\citeResults).
The minimum support avoids overly infrequent (i.e., non-generalizable) rules from being created,
 and the minimum confidence prevents the creation of imprecise rules.
Note that no rule (with \formatItemName{NotFaulty} as rule consequent) can reach a higher support than 50\% after the SMOTE pre-processing.
After computing the rules, we removed redundant ones using the corresponding function from the \textit{apriori} package.
We then sorted the remaining rules descending by their confidence.

Using these rules, we created two classifiers to identify low-fault-risk (LFR) methods.
They differ in the number of comprised rules.
 The strict classifier uses the top \textit{n} rules until the share of faulty methods in all methods (of the training set) exceeds 2.5\% in the LFR methods (of the training set).
 The more lenient classifier uses the top \textit{n} rules until the share exceeds 5\% in the LFR methods.
(Example: We applied the top one rule to the training set,
 then applied the next rule, \dots, until the matched methods in the training set contained 2.5\% out of all faults.)
Figure~\ref{Fig:RuleNumberInfluence} presents
 how an increase in the number of selected rules affects the proportion of LFR methods
 and the share of faulty methods that they contain.
For \rqNumFaultsInLfr and \rqNumSizeLfr, the classifiers were computed for each fold of each project.
For \rqNumCrossPrj, the classifiers were computed once for each project.

\definecolor{Color_RNI_LFR}{RGB}{2, 103, 161}
\definecolor{Color_RNI_FS}{RGB}{250, 111, 111}

\begin{figure}
	\centering
	\includegraphics[width=0.6\linewidth]{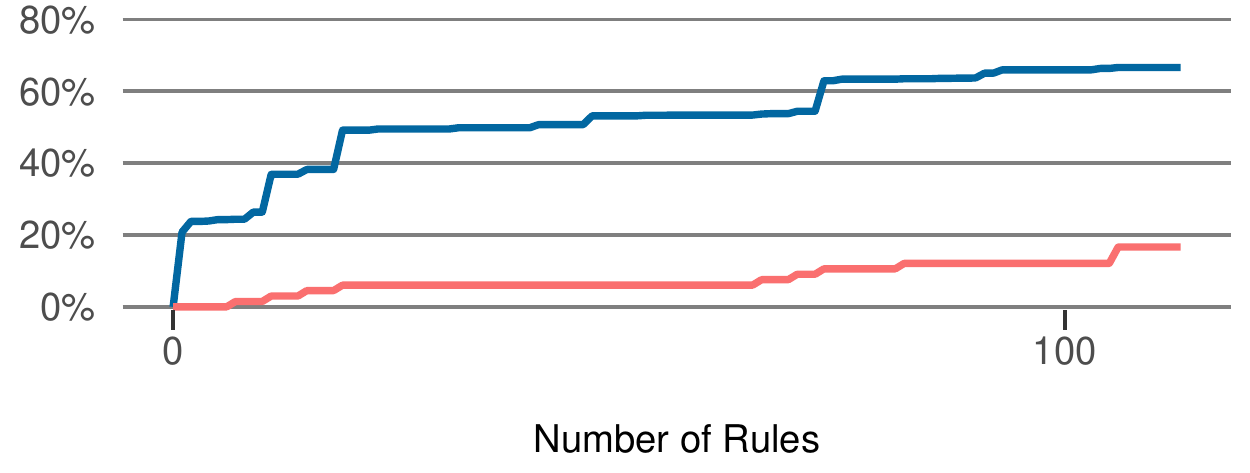}
	\caption[Influence of the number of selected rules (\textit{Lang}).]{
					 Influence of the number of selected rules (\textit{Lang}).
					 \formatCaptionDetails{The number of rules influences
					 the \coloredLine{Color_RNI_LFR}~proportion of low-fault-risk (LFR) methods
					 and the \coloredLine{Color_RNI_FS}~share of faulty methods in LFR out of all faulty methods.}
					}
	\label{Fig:RuleNumberInfluence}
\end{figure}

\vspace{\halfBaselineskip}
To answer \textbf{\rqNumFaultsInLfr,}
 we used 10-fold cross-validation to evaluate the classifiers separately for each project.
We computed the number and proportion of methods that were classified as ``low-fault-risk'' but contained a fault ($\approx$~false positives).
For the sake of completeness, we also computed precision and recall;
 although, we believe that the recall is of lesser importance for our purpose.
This is because we do not want to predict \textit{all} methods that do not contain any faults in the dataset;
 we only want to identify those methods that we can say, \textit{with high certainty}, contain hardly any faults.

As the dataset is imbalanced with faulty methods in the minority,
 the proportion of faults in low-fault-risk methods might not be sufficient to assess the classifiers
 (SMOTE was applied only to the training set).
Therefore, we further computed the \textit{\fdr},
 which describes how much less likely the LFR methods contain a fault.
For example, if 40\% of all methods are classified as ``low fault risk'' and contain 10\% of all faults, the \fdrShort is 4.
It can also be read as: 40\% of all methods contain only one fourth of the expected faults.
We mathematically define the \fdrLong based on methods as
\vspace{\miniSkip}
\begin{center}
 $\frac{\text{proportion of LFR methods out of all methods}}{\text{proportion of faulty LFR methods out of all faulty methods}}$
\end{center}
\vspace{\miniSkip}
 and based on SLOC as
\vspace{\miniSkip}
\begin{center}
 $\frac{\text{proportion of SLOC in LFR methods out of all SLOC}}{\text{proportion of faulty LFR methods out of all faulty methods}}$.
\end{center}
\vspace{\miniSkip}
For both classifiers (strict variant with 2.5\%, lenient variant with 5\%),
 we present the metrics for each project and the resulting median.

\vspace{\halfBaselineskip}
To answer \textbf{\rqNumSizeLfr,}
 we assessed how common methods classified as ``low fault risk'' are.
For each project, we computed the absolute number of low-fault-risk methods, their proportion out of all methods,
 and their extent by considering their SLOC.
\textit{LFR SLOC} corresponds to the sum of SLOC of all low-fault-risk methods.
The proportion of LFR SLOC is computed out of all SLOC of the project.

\vspace{\halfBaselineskip}
To answer \textbf{\rqNumCrossPrj,}
 we computed the association rules for each project with the methods of the other five projects as training data.
Like in \rqNumFaultsInLfr and \rqNumSizeLfr,
 we determined the number of used top \textit{n} rules with the same thresholds (2.5\% and 5\%).
To allow a comparison with the within-project classifiers,
 we computed the same metrics like in \rqNumFaultsInLfr and \rqNumSizeLfr.

\subsection{Results}
This section presents the results to the research questions.
The data to reproduce the results is available at~\citeResults.

\begin{table}
	\centering
	\footnotesize
	\addtolength{\leftskip}{-2cm}
  \addtolength{\rightskip}{-2cm}
	\caption{\rqNumFaultsInLfr, \rqNumSizeLfr: Evaluation of within-project IDP to identify low-fault-risk (LFR) methods.}
	\begin{tabular}
	{lR{0.4cm}R{0.65cm}R{0.9cm}R{0.75cm}R{0.8cm}R{0.75cm}R{0.85cm}R{0.75cm}R{1.55cm}R{1.15cm}R{0.95cm}}
		\toprule

Project				& \multicolumn{2}{r}{Faults in LFR} & \multicolumn{2}{r}{LFR methods} & \multicolumn{2}{r}{LFR methods} & \multicolumn{2}{r}{LFR SLOC}	& LFR methods		& \multicolumn{2}{r}{\fdr} \\
							& &	& & & & & & & contain	\dots\% & & \\
							& \# & \% 													& Prec. & Rec.                    & \# & \% 												& \# & \% 											& of all faults	  & (methods) & (SLOC) \\

\midrule

\rowcolor{\tableRowBgColor}
\multicolumn{12}{l}{\small{\textit{\textbf{Within-project IDP, 10-fold:} min. support = \minSuppVal, min. confidence = \minConfVal, rules until fault share in training set = \textbf{2.5\%}}}}\\
\midrule
		Chart & 4 & 0.1\% &  99.9\% & 44.1\% & 2,995 & 43.9\% & 11,228 & 15.8\% & 10.3\% & 4.3 & 1.5 \\
		Closure & 6 & 0.2\% &  99.8\% & 29.2\% & 3,759 & 28.9\% & 15,497 & 10.5\% & 4.1\% & 7.1 & 2.6 \\
		Lang & 3 & 0.5\% &  99.5\% & 29.6\% & 576 & 28.6\% & 2,242 & 13.8\% & 4.1\% & 7.0 & 3.4 \\
		Math & 2 & 1.1\% & 98.9\% & 18.4\% & 190 & 16.5\% & 570 & 4.8\% & 1.5\% & 10.9 & 3.1 \\
		Mockito & 5 & 0.6\% & 99.4\% & 35.1\% & 875 & 34.4\% & 6,128 & 25.1\% & 7.8\% & 4.4 & 3.2 \\
		Time & 8 & 0.1\% &  99.9\% & 80.4\% & 8,063 & 80.2\% & 62,063 & 78.1\% & 17.8\% & 4.5 & 4.4 \\
		\textit{Median} & \textit{} & \textit{0.3\%} & \textit{ 99.7\%} & \textit{32.3\%} & \textit{} & \textit{31.7\%} & \textit{} & \textit{14.8\%} & \textit{6.0\%} & \textit{5.7} & \textit{3.2} \\

\midrule

\rowcolor{\tableRowBgColor}
\multicolumn{12}{l}{\small{\textit{\textbf{Within-project IDP, 10-fold:} min. support = \minSuppVal, min. confidence = \minConfVal, rules until fault share in training set = \textbf{5\%}}}}\\
\midrule
		Chart & 4 & 0.1\% &  99.9\% & 44.8\% & 3,040 & 44.6\% & 11,563 & 16.3\% & 10.3\% & 4.3 & 1.6 \\
		Closure & 15 & 0.3\% &  99.7\% & 41.8\% & 5,385 & 41.5\% & 25,981 & 17.6\% & 10.1\% & 4.1 & 1.7 \\
		Lang & 6 & 0.7\% & 99.3\% & 45.0\% & 879 & 43.7\% & 3,630 & 22.3\% & 8.2\% & 5.3 & 2.7 \\
		Math & 7 & 2.7\% & 97.3\% & 24.3\% & 255 & 22.1\% & 878 & 7.3\% & 5.3\% & 4.2 & 1.4 \\
		Mockito & 6 & 0.5\% &  99.5\% & 47.8\% & 1,189 & 46.8\% & 8,260 & 33.8\% & 9.4\% & 5.0 & 3.6 \\
		Time & 9 & 0.1\% &  99.9\% & 82.8\% & 8,298 & 82.5\% & 63,333 & 79.7\% & 20.0\% & 4.1 & 4.0 \\
		\textit{Median} & \textit{} & \textit{0.4\%} & \textit{ 99.6\%} & \textit{44.9\%} & \textit{} & \textit{44.1\%} & \textit{} & \textit{20.0\%} & \textit{ 9.8\%} & \textit{4.3} & \textit{2.2} \\

\bottomrule
	\end{tabular}
	\label{Tbl:Results_RQ2_RQ3}
\end{table}

\begin{table}
	\centering
	\caption{Top three association rules for \textit{Lang} \formatCaptionDetails{(within-project, fold 1).}}
	\begin{tabular}{l>{\raggedright}p{9.7cm}rr}
		\toprule
			\# & Rule		 		& Support & Confidence \\
		\midrule
		
		1 & \small{
		 \{ \formatItemNameSmall{UniqueVariableIdentifiersLessThan2},
     \formatItemNameSmall{NoMethodInvocations} \}
		 \mbox{$\Rightarrow$~\{ \formatItemNameSmall{NotFaulty} \}}} & 10.98\%  & 100.00\% \\
		
		2 & \small{
		 \{ \formatItemNameSmall{SlocLessThan4},
     \formatItemNameSmall{NoMethodInvocations},
		 \formatItemNameSmall{NoArithmeticOperations} \}
		 \mbox{$\Rightarrow$~\{ \formatItemNameSmall{NotFaulty} \}}} & 10.98\%  & 100.00\% \\
		
		3 & \small{
		 \{ \formatItemNameSmall{SlocLessThan4},
     \formatItemNameSmall{NoMethodInvocations},
		 \formatItemNameSmall{NoCastExpressions} \}
		 \mbox{$\Rightarrow$~\{ \formatItemNameSmall{NotFaulty} \}}} & 10.60\%  & 100.00\% \\
				
		\bottomrule
	\end{tabular}
	\label{Tbl:TopRules}
\end{table}

\vspace{\halfBaselineskip}
\textbf{\rqTextFaultsInLfr}
Table~\ref{Tbl:Results_RQ2_RQ3} presents the results.
The methods classified to have low fault risk (LFR) by the stricter classifier, 
 which allows a maximum fault share of 2.5\% in the LFR methods in the (balanced) training data,
 contain between 2 and 8 faulty methods per project.
The more lenient classifier,
 which allows a maximum fault share of 5\%,
 classified between 4 and 15 faulty methods as LFR.
The median proportion of faulty methods in LFR methods is 0.3\% resp. 0.4\%.

The \fdrLong for the stricter classifier ranges
 between 4.3 and 10.9 (median: 5.7) when considering methods
 and between 1.5 and 4.4 (median: 3.2) when considering SLOC.
In the project \textit{Lang}, 28.6\% of all methods with 13.8\% of the SLOC are classified as LFR and contain 4.1\% of all faults,
 thus, the \fdrShort is 7.0 (SLOC-based: 3.4).
The \fdrShort never falls below 1 for both classifiers.

\formatResultBox{
 IDP can identify methods with low fault risk.
 On average, only 0.3\% of the methods classified as ``low fault risk'' by the strict classifier are faulty.
 The identified LFR methods are, on average, 5.7 times less likely to contain a fault than an arbitrary method in the dataset.}

\vspace{\halfBaselineskip}
Table~\ref{Tbl:TopRules} exemplarily presents the top three rules for \textit{Lang}.
Methods that work with fewer than two variables and do not invoke any methods
 as well as short methods without arithmetic operations, cast expressions, and method invocations are highly unlikely to contain a fault.

\vspace{\halfBaselineskip}
\textbf{\rqTextSizeLfr}
Table~\ref{Tbl:Results_RQ2_RQ3} presents the results.
The stricter classifier classified between 16.5\% and 80.2\% of the methods as LFR (median: 31.7\%, mean: 38.8\%),
 the more lenient classifier matched between 22.1\% and 82.5\% of the methods (median: 44.1\%, mean: 46.9\%).
The median of the comprised SLOC in LFR methods is 14.8\% (mean: 24.7\%)
 respectively 20.0\% (mean: 29.5\%).

\formatResultBox{
 Using within-project IDP, on average, 32--44\% of the methods,
  comprising about 15--20\% of the SLOC,
  can be assigned a lower importance during testing.
}

\formatResultBox{
  In the best case, when ignoring 16.5\% of the methods (4.8\% of the SLOC),
   it is still possible to catch 98.5\% of the faults (\textit{Math}).
}

\vspace{\halfBaselineskip}
\textbf{\rqTextCrossPrj}
Table~\ref{Tbl:Results_RQ4} presents the results for the cross-project prediction
 with training data from the respective other projects.
Compared to the results of the within-project prediction, except for \textit{Math},
 the number of faults in LFR methods decreased or stayed the same in all projects for both classifier variants.
While the median proportion of faults in LFR methods slightly decreased,
the proportion of LFR methods also decreased in all projects except \textit{Math}.
The median proportion of LFR methods is 23.3\% (SLOC: 8.1\%) for the stricter classifier
 and 26.3\% (SLOC: 12.6\%) for the more lenient classifier.

The \fdr improved compared to the within-project prediction
 for both the method and SLOC level in both classifier variants:
For the stricter classifier, the median of the method-based \fdrShort is 10.9 (+5.2);
 the median of the SLOC-based \fdrShort is 3.9 (+0.7).
Figures~\ref{Fig:ComparisonWithinAndCrossProject} illustrates the \fdr for both
 within-project (\rqNumFaultsInLfr, \rqNumSizeLfr)
 and cross-project (\rqNumCrossPrj) prediction.

\definecolor{Color_RQ5_IDP_SP_2p5}{RGB}{237, 248, 233} 
\definecolor{Color_RQ5_IDP_SP_5p0}{RGB}{186, 228, 179} 
\definecolor{Color_RQ5_IDP_CP_2p5}{RGB}{107, 174, 214} 
\definecolor{Color_RQ5_IDP_CP_5p0}{RGB}{ 33, 113, 181} 

\begin{figure}
	\centering
	\includegraphics[width=1.0\linewidth,clip]{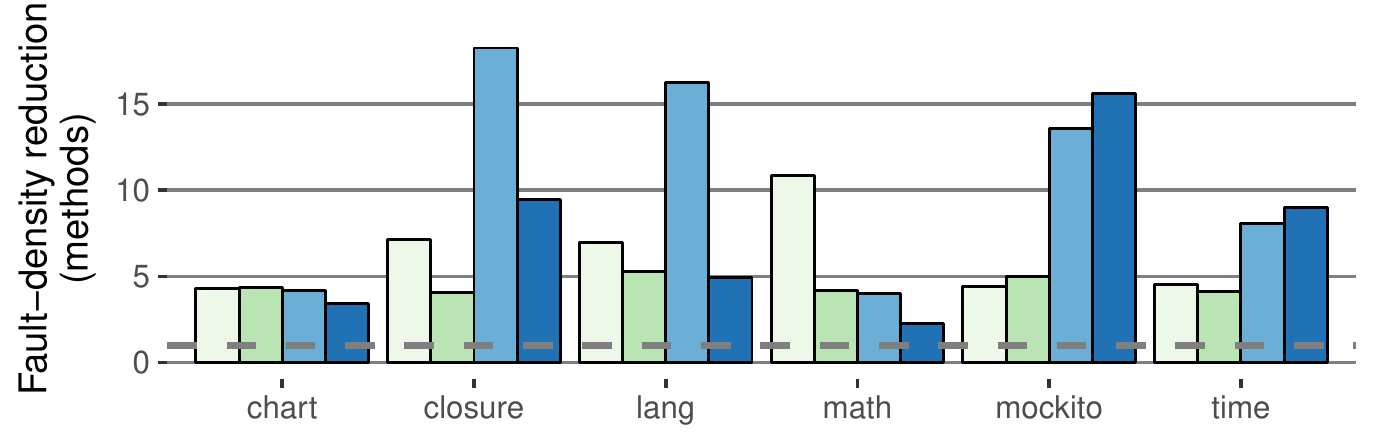}
	\caption[Comparison of the within- and cross-project IDP classifiers.]{
		Comparison of the IDP within-project (\coloredSquare{Color_RQ5_IDP_SP_2p5}~2.5\%, \coloredSquare{Color_RQ5_IDP_SP_5p0}~5.0\%)
		 with the IDP cross-project (\coloredSquare{Color_RQ5_IDP_CP_2p5}~2.5\%, \coloredSquare{Color_RQ5_IDP_CP_5p0}~5.0\%) classifiers
		 (method-based).
		\formatCaptionDetails{
		 The \fdr expresses how much less likely a LFR method contains a fault (definition in~\ref{Sec:Procedure}).
		 Higher values are better.
		 (Example: If 40\% of the methods are LFR and contain 5\% of all faults, the \fdrShort is 8.)
		 The dashed line is at one; no value falls below.
		}
	}
	\label{Fig:ComparisonWithinAndCrossProject}
\end{figure}

\formatResultBox{
 Using cross-project IDP, on average, 23--26\% of the methods,
  comprising about 8--13\% of the SLOC,
  can be classified as ``low fault risk''.
 The methods classified by the stricter classifier contain, on average, less than one eleventh of the expected faults.}

\begin{table}
	\centering
	\footnotesize
  \addtolength{\leftskip}{-2cm}
  \addtolength{\rightskip}{-2cm}
	\caption{\rqNumCrossPrj: Evaluation of cross-project IDP.}
	\begin{tabular}
	{lR{0.4cm}R{0.65cm}R{0.9cm}R{0.75cm}R{0.8cm}R{0.75cm}R{0.85cm}R{0.75cm}R{1.55cm}R{1.15cm}R{0.95cm}}
		\toprule
		
Project				& \multicolumn{2}{r}{Faults in LFR} & \multicolumn{2}{r}{LFR methods} & \multicolumn{2}{r}{LFR methods} & \multicolumn{2}{r}{LFR SLOC}	& LFR methods		& \multicolumn{2}{r}{\fdr} \\
							& &	& & & & & & & contain	\dots\% & & \\
							& \# & \% 													& Prec. & Rec.                    & \# & \% 												& \# & \% 											& of all faults	  & (methods) & (SLOC) \\

		\midrule
		
\rowcolor{\tableRowBgColor}
\multicolumn{12}{l}{\small{\textit{\textbf{Cross-project IDP:} min. support = \minSuppVal, min. confidence = \minConfVal, rules until fault share in training set = \textbf{2.5\%}}}}\\
\midrule
		Chart & 3 & 0.1\% &  99.9\% & 32.1\% & 2,182 & 32.0\% & 7,434 & 10.5\% & 7.7\% & 4.2 & 1.4 \\
		Closure & 2 & 0.1\% &  99.9\% & 25.0\% & 3,207 & 24.7\% & 11,584 & 7.9\% & 1.4\% & 18.3 & 5.8 \\
		Lang & 1 & 0.2\% &  99.8\% & 23.1\% & 449 & 22.3\% & 1,357 & 8.3\% & 1.4\% & 16.3 & 6.1 \\
		Math & 8 & 2.9\% & 97.1\% & 26.6\% & 280 & 24.3\% & 1,129 & 9.4\% & 6.1\% & 4.0 & 1.6 \\
		Mockito & 1 & 0.2\% &  99.8\% & 21.7\% & 539 & 21.2\% & 1,698 & 6.9\% & 1.6\% & 13.6 & 4.4 \\
		Time & 1 & 0.1\% &  99.9\% & 18.4\% & 1,845 & 18.3\% & 5,807 & 7.3\% & 2.2\% & 8.3 & 3.3 \\
		\textit{Median} & \textit{} & \textit{0.2\%} & \textit{ 99.8\%} & \textit{24.0\%} & \textit{} & \textit{23.3\%} & \textit{} & \textit{8.1\%} & \textit{1.9\%} & \textit{10.9} & \textit{3.9} \\

\midrule
\rowcolor{\tableRowBgColor}
\multicolumn{12}{l}{\small{\textit{\textbf{Cross-project IDP:} min. support = \minSuppVal, min. confidence = \minConfVal, rules until fault share in training set = \textbf{5\%}}}}\\
\midrule
		Chart & 4 & 0.2\% &  99.8\% & 35.5\% & 2,411 & 35.4\% & 9,363 & 13.2\% & 10.3\% & 3.4 & 1.3 \\
		Closure & 4 & 0.1\% &  99.9\% & 25.9\% & 3,327 & 25.6\% & 15,583 & 10.6\% & 2.7\% & 9.5 & 3.9 \\
		Lang & 4 & 0.7\% & 99.3\% & 27.7\% & 542 & 26.9\% & 1,959 & 12.0\% & 5.5\% & 4.9 & 2.2 \\
		Math & 18 & 5.1\% & 94.9\% & 32.9\% & 354 & 30.7\% & 1,634 & 13.7\% & 13.6\% & 2.2 & 1.0 \\
		Mockito & 1 & 0.2\% &  99.8\% & 25.0\% & 620 & 24.4\% & 3,495 & 14.3\% & 1.6\% & 15.6 & 9.1 \\
		Time & 1 & 0.0\% & 100.0\% & 20.0\% & 2,007 & 20.0\% & 7,552 &  9.5\% & 2.2\% & 9.0 & 4.3 \\
		\textit{Median} & \textit{} & \textit{0.2\%} & \textit{ 99.8\%} & \textit{26.8\%} & \textit{} & \textit{26.3\%} & \textit{} & \textit{12.6\%} & \textit{4.1\%} & \textit{6.9} & \textit{3.1} \\

		\bottomrule
	\end{tabular}
	\label{Tbl:Results_RQ4}
\end{table}

\section{Discussion}
\label{Sec:Discussion}

The results of our empirical study show that only very few low-fault-risk methods actually contain a fault,
 and thus, they indicate that IDP can successfully identify methods that are not fault-prone.
On average, 31.7\% of the methods (14.8\% of the SLOC) matched by the strict classifier contain only 6.0\% of all faults,
 resulting in a considerable \fdr for the matched methods.
In any case, low-fault-risk methods are less fault-prone than other methods,
 (\fdr is higher than one in all projects);
 based on methods, LFR methods are at least twice less likely to contain a fault.
For the stricter classifier, the extent of the matched methods, which could be deferred in testing, is between 5\% and 78\% of the SLOC of the respective project.
The more lenient classifier matches more methods and SLOC at the cost of a higher fault proportion,
 but still achieves satisfactory \fdr values.
This shows that the balance between fault risk and matched extent can be influenced
 by the number of considered rules to reflect the priorities of a software project.

Interestingly, the cross-project IDP classifier, which is trained on data from the respective other five projects,
 exhibits a higher precision than the within-project IDP classifier.
Except for the \textit{Math} project, the LFR methods contain fewer faulty methods in the cross-project prediction scenario.
This is in line with the method-based \fdrLong of the strict classifier, which is in four of six cases better in the cross-project scenario (SLOC-based: three of six cases).
However, the proportion of matched methods decreased compared to the within-project prediction in most projects.
Accordingly, the cross-project results suggest that a larger, more diversified training set identifies LFR methods more conservatively,
 resulting in a higher precision and lower matching extent.

\textit{Math} is the only project in which IDP within-project prediction outperformed IDP cross-project prediction.
This project contains many methods with mathematical computations expressed by arithmetic operations,
 which are often wrapped in loops or conditions;
 most of the faults are located in these methods.
Therefore, the within-project classifiers used few, very precise rules for the identification of LFR methods.

To sum up, our results show that the IDP approach can be used to identify methods
 that are, due to the ``triviality'' of their code,
 less likely to contain any faults.
Hence, these methods require less focus during quality-assurance activities.
Depending on the criticality of the system and the risk one is willing to take,
 the development of tests for these methods can be deferred or even omitted in case of insufficient available test resources.
The results suggest that IDP is also applicable in cross-project prediction scenarios,
 indicating that characteristics of low-fault-risk methods differ less between projects than characteristics of faulty methods do.
Therefore, IDP can be used in (new) projects with no (precise) historical fault data to prioritize the code to be tested.

\subsection{Limitations}

A limitation of IDP is that even low-fault-risk methods can contain faults.
An inspection of faulty methods incorrectly classified to have a low fault risk showed
 that some faults were fixed by only adding further statements (e.g., to handle special cases).
This means that a method can be faulty even if the existing code as such is not faulty (due to missing code).
Further imaginable examples for faulty low-fault-risk methods are
 simple getters that return the wrong variable,
 or empty methods that are unintentionally empty.
Therefore, while these methods are much less fault-prone, it cannot be assumed that they never contain any fault.
Consequently, excluding low-fault-risk methods from testing and other QA activities carries a risk that needs to be kept in mind.

\subsection{Relation to Defect Prediction}

As discussed in detail in Section~\ref{Sec:Introduction},
 IDP presents another view on defect prediction.
The focus of IDP on low-fault-risk methods allows optimizing towards precision,
 while recall is less important.
Therefore, a precision-and-recall comparison of our study results with method-level defect prediction studies from other papers,
 such as \cite{giger2012method} or~\cite{hata2012bug},
 would lead to a performance comparison of the used metrics or classifiers,
 which is not what differentiates IDP from traditional defect prediction.
\subsection{Threats to Validity}
Next, we discuss the threats to internal and external validity.

\subsubsection{Threats to Internal Validity}

The learning and evaluation was performed on information extracted from Defects4J~\cite{just2014defects4j}.
Therefore, the quality of our data depends on the quality of Defects4J.
Common problems for defect datasets created by analyzing changes in commits that reference a bug ticket in an issue tracking system are as follows.
First, commits that fix a fault but do not reference a ticket in the commit message cannot be detected~\cite{bachmann2010missing}.
 Consequently, the set of commits that reference a bug fix may not be a fair representation of all faults~\cite{bird2009fair, d2012evaluating, giger2012method}.  
Second, bug tickets in the issue tracker may not always represent faults and vice versa.
Herzig et al. pointed out that a significant amount of tickets in the issue trackers of open-source projects is misclassified~\cite{herzig2013misclassification}.
Therefore, it is possible that not all bug-fix commits were spotted.
Third, faults may not have been detected or fixed yet.
In general, it is not possible to prove that a method does not contain any faults.
Fourth, a commit may contain changes (such as refactorings) that are not related to the bug fix,
 but this problem does not affect the Defects4J dataset due to the authors' manual inspection.
These threats are present in nearly all defect prediction studies, especially in those operating at the method level.
Defect prediction models were found to be resistant to such kind of noise to a certain extent~\cite{kim2011dealing}.

Defects4J contains only faults that are reproducible and can be precisely mapped to methods;
 therefore, faulty methods may be under-approximated.
In contrast, other datasets created without manual post-processing tend to over-approximate faults.
To mitigate this threat, we replicated our IDP evaluation with two study objects used in \cite{giger2012method} by Giger et al.
The observed results were similar to our study.

\subsubsection{Threats to External Validity}

The empirical study was performed with six mature open-source projects written in Java.
The projects are libraries and their results may not be applicable to other application types,
 e.g., large industrial systems with user interfaces.
The results may also not be transferable to projects of other languages, for the following reasons:
First, Java is a strongly typed language that provides type safety.
It is unclear if the IDP approach works for languages without type safety,
 because it could be that even simple methods in such languages exhibit a considerable amount of faults.
Second, in case the approach as such is applicable to other languages,
 the collected metrics and the low-fault-risk classifier need to be validated and adjusted.
Other languages may use language constructs in a different way or use constructs that do not exist in Java.
For example, a classifier for the C language should take constructs such as \texttt{GOTO}s and the use of pointer arithmetic into consideration.
Furthermore, the projects in the dataset (published in 2014) did not contain code with lambda expressions introduced in Java~8.\footnote{\url{http://www.oracle.com/technetwork/articles/java/architect-lambdas-part1-2080972.html}}
Therefore, in newer projects that make use of lambda expressions,
 the presence of lambdas should be taken into consideration when classifying methods.
Consequently, further studies are necessary to determine whether the results are generalizable.

As done in most defect prediction studies, we treated all faults as equal and did not consider their importance.
In reality, not all faults have the same importance, because some cause higher failure follow-up costs than others.

\section{Conclusion}
\label{Sec:Conclusion}
Developer teams often face the problem scarce test resources and need therefore to prioritize their testing efforts
 (e.g., when writing new automated unit tests).
Defect prediction can support developers in this activity.
In this paper, we propose an inverse view on defect prediction (IDP) to identify methods
 that are so ``trivial'' that they contain hardly any faults.
We study how unerringly such low-fault-risk methods can be identified,
 how common they are,
 and whether the proposed approach is applicable for cross-project predictions.

We show that IDP using association rule mining on code metrics can successfully identify low-fault-risk methods.
The identified methods contain considerably fewer faults than the average code
 and can provide a savings potential for QA activities.
Depending on the parameters, a lower priority for QA can be assigned on average to 31.7\% resp. 44.1\% of the methods,
 amounting to 14.8\% resp. 20.0\% of the SLOC.
While cross-project defect prediction is a challenging task~\cite{he2012investigation, zimmermann2009cross},
 our results suggest that the IDP approach can be applied in a cross-project prediction scenario at the method level.
In other words, an IDP classifier trained on one or more (Java open-source) projects can successfully
 identify low-fault-risk methods in other Java projects for which no---or no precise---fault data exists.

For future work, we want to replicate this study with closed-source projects, projects of other application types,
 and projects in other programming languages.
It is also of interest to investigate which metrics and classifiers are most effective for the IDP purpose
 and whether they differ from the ones used in traditional defect prediction.
Moreover, we plan to study
 whether code coverage of low-fault-risk methods differs from code coverage of other methods.
If guidelines to meet a certain code coverage level are set by the management,
 unmotivated testers may add tests for low-fault-risk methods first because it might be easier to write tests for those methods.
Consequently, more complex methods with a higher fault risk may remain untested once the target coverage is achieved.
Therefore, we want to investigate whether this is a problem in industry
 and whether it can be addressed with an adjusted code-coverage computation,
 which takes low-fault-risk methods into account.

\section*{Acknowledgment}
This work was partially funded by the German Federal Ministry of Education and Research (BMBF), grant ``SOFIE, 01IS18012A''.
The responsibility for this article lies with the authors.
We thank
 Nils G\"ode
 and Florian Dei{\ss}enb\"ock
 for their valuable feedback.

\balance

\bibliography{./literature/bibtex} 

\end{document}